\documentclass[pre,preprint,showpacs,amsmath,amssymb,superscriptaddress,letterpaper]{revtex4}

\usepackage{graphicx}% Include figure files
\usepackage{subfigure}

\begin{document}

\title{Optimal Stimulus and Noise Distributions for Information Transmission via Suprathreshold Stochastic Resonance}

\author{Mark D. McDonnell}
 \email{mmcdonne@eleceng.adelaide.edu.au (all correspondence)}
\affiliation{ School of Electrical and Electronic Engineering \&
Centre for Biomedical Engineering, The University of Adelaide, SA
5005, Australia }
\author{Nigel G. Stocks}
\affiliation{School of Engineering, The University of Warwick,
Coventry CV4 7AL, United Kingdom }
\author{Derek Abbott}
\affiliation{ School of Electrical and Electronic Engineering \&
Centre for Biomedical Engineering, The University of Adelaide, SA
5005, Australia }

\date{\today}

\begin{abstract}
Suprathreshold stochastic resonance (SSR) is a form of noise
enhanced signal transmission that occurs in a parallel array of
independently noisy identical threshold nonlinearities, including
model neurons. Unlike most forms of stochastic resonance, the output
response to suprathreshold random input signals of {\em arbitrary
magnitude} is improved by the presence of even small amounts of
noise. In this paper the information transmission performance of SSR
in the limit of a large array size is considered. Using a
relationship between Shannon's mutual information and Fisher
information, a sufficient condition for optimality, i.e.~channel
capacity, is derived. It is shown that capacity is achieved when the
signal distribution is Jeffrey's prior, as formed from the noise
distribution, or when the noise distribution depends on the signal
distribution via a cosine relationship. These results provide
theoretical verification and justification for previous work in both
computational neuroscience and electronics.
\end{abstract}


\maketitle

\section{Introduction}\label{s:intro}

The term `stochastic resonance' describes the situation where a
system's response to some signal is optimized by the presence of
random noise, rather than its absence. It occurs in a wide variety
of nonlinear physical~\cite{Gammaitoni.Jan98} and
biological~\cite{Moss.04} systems.

In many of the systems and models in which stochastic resonance (SR)
has been observed, the essential nonlinearity is a single static
threshold,
e.g.~\cite{Wiesenfeld.94,Bulsara.96,Chapeau.May96,Greenwood.99}. It
is generally thought that SR cannot occur in such systems for
suprathreshold signals, meaning that the amplitude of the input
signal needs to be restricted to values smaller than the amplitude
of the threshold for SR to occur~\cite{DeWeese.95}.

\begin{figure}[htbp]
\centering\includegraphics[scale=0.5]{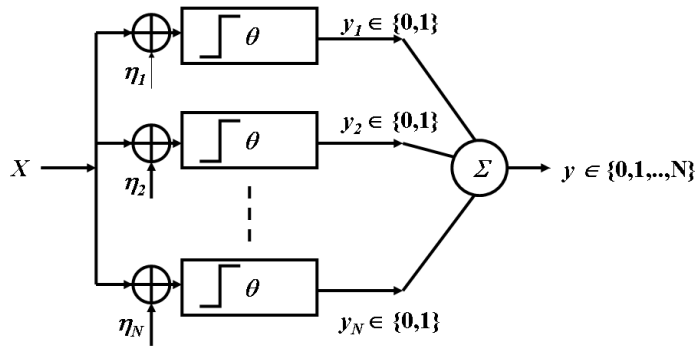}
\caption[]{The SSR model consists of $N$ parallel threshold devices,
each with the same threshold value, $\theta$. The common input
signal is a continuously valued random signal, $X$, consisting of a
sequence of discrete time uncorrelated samples. Each device receives
independently noisy versions of $X$. The noise signals, $\eta_i$,
are {\it iid} additive random signals that are independent of $X$.
The output from the $i$--th device, $y_i$, is unity if
$X+\eta_i>\theta$ and zero otherwise. The overall output, $y$, is
the sum of the individual outputs, $y_i$.} \label{f:4:SSR_Array}
\end{figure}

However, the 1999 discovery of a novel form of SR---known as {\em
suprathreshold stochastic resonance} (SSR)---showed that this is not
always true~\cite{Stocks.Mar2000}. SSR occurs in an array of
identical threshold nonlinearities, each of which are subject to
independently random additive noise. We refer to this array as the
{\em SSR model}---see Fig.~\ref{f:4:SSR_Array}. In this model SR
occurs regardless of whether the input signal is entirely
subthreshold or not. Furthermore, SSR occurs even for very large
input SNRs. This is a further difference to conventional SR, for
which the signal is required to be weak compared to the noise.

SSR is a form of aperiodic stochastic
resonance~\cite{Collins.95,Bulsara.96,Morse.00} that was first shown
to occur by calculating Shannon's average mutual information for the
SSR model~\cite{Stocks.Mar2000}.  It was subsequently found that the
performance achievable via SSR is maximized when all threshold
values are set to the signal mean~\cite{Stocks.01b}, and that for
sufficiently small input SNRs, modifying the thresholds in the model
cannot improve information transfer~\cite{McDonnell.06}.

The SSR model was originally motivated as a model for parallel
sensory neurons, such as those synapsing with hair cells in the
inner ear~\cite{Stocks.02}. Although the basic SSR model is
non-dynamical, and does not model the many complexities of real
neurons, each threshold nonlinearity is equivalent to a
Pitts-McCulloch neuron model, and encapsulates the neural coding
properties we are interested in---i.e. the generation of action
potentials in response to a noisy aperiodic random stimulus. The
small input SNRs we focus on are biologically
relevant~\cite{Bialek.93}, particularly so for hair cells, which are
subject to substantial Brownian motion~\cite{Lindner.05}. This leads
to much randomness in the release of neurotransmitters at synapses
with afferent neurons leading to the cochlear nucleus.

Further justification of the SSR model's relevance to neural coding
is discussed in~\cite{Hoch.03a,Durrant.06,Nikitin.07}, and by
extensions of the model to include more biologically realistic
neural features. For example, the parallel array has been modified
to consist of parallel FitzHugh-Nagumo neuron
models~\cite{Stocks.01c}, leaky integrate-and-fire neuron
models~\cite{Hoch.03a,Durrant.06} and Hodgkin-Huxley
models~\cite{Hoch.03a}, and for the case of signal-dependent
(multiplicative) noise~\cite{Nikitin.07}. In all cases the same
qualitative results as for the simple threshold model were obtained.
The SSR effect has also led to a proposal for improving the
performance of cochlear implants for suprathreshold
stimuli~\cite{Stocks.02}, based on the idea that the natural
randomness present in functioning cochlear hair cells is missing in
patients requiring implants~\cite{Morse.07}.

The purpose of this paper is to analyze, in a {\em general} manner,
the information theoretic upper limits of performance of the SSR
model. This requires allowing the array size, $N$, to approach
infinity. Previous work has discussed the scaling of the mutual
information through the SSR model with $N$ for specific cases, and
found conditions for which the maximum mutual information---i.e.
channel capacity---occurs~\cite{Stocks.01a,Stocks.01b,Hoch.03a}. In
a neural coding context, the question of `what is the optimal
stimulus distribution?' for a given noise distribution is discussed
numerically for the SSR model in~\cite{Hoch.03a}.

In Sec.~\ref{c:5capacity}, we significantly extend the results
in~\cite{Stocks.01a,Stocks.01b,Hoch.03a}, by showing that the mutual
information and output entropy can both be written in terms of
simple relative entropy expressions---see Eqs.~(\ref{new})
and~(\ref{DP_S}). This leads to a very general sufficient condition,
Eq.~(\ref{sufficient}), for achieving capacity in the large $N$
regime that can be achieved either by optimizing the signal
distribution for a given noise distribution, or optimizing the noise
for a given signal. Given the neuroscience motivation for studying
the SSR model, this result is potentially highly significant in
computational neuroscience, where both optimal stimulus
distributions, and optimal tuning curves are often
considered~\cite{Hoch.03a,Brunel.98}.

Furthermore, the optimal signal for the special case of uniform
noise is shown to be the arcsine distribution (a special case of the
Beta distribution), which has a relatively large variance and is
bimodal. This result provides theoretical justification for a
proposed heuristic method for analog-to-digital conversion based on
the SSR model~\cite{Nguyen.06}. In this method, the input signal is
transformed so that it has a large variance and is bimodal.

As a means of verification of our theory, in Sec.~\ref{c:compare}
our general results are compared to the specific capacity results
contained in~\cite{Stocks.01a,Stocks.01b,Hoch.03a}. This leads us to
find and justify improvements to these previous results.

Before we proceed however, the remainder of this section outlines
our notation, describes the SSR model, and derives some important
results that we utilize.

\subsection{Information Theoretic Definitions}

Recent work using the SSR model has described performance using
measures other than mutual
information~\cite{McDonnell.02MEJ,McDonnell.02FNL,Rousseau.03b,Rousseau.04,McDonnell.05,Wang.05}.
However, in line with much theoretical neuroscience
research~\cite{Bialek.93}, here we use the information theoretic
viewpoint where the SSR model can be considered to be a
communications channel~\cite{Stocks.Mar2000}.

Throughout, we denote the probability mass function (PMF) of a
discrete random variable, $\alpha$, as $P_\alpha(\cdot)$, the
probability density function (PDF) of a continuous random variable,
$\beta$, as $f_\beta(\cdot)$, and the cumulative distribution
function (CDF) of $\beta$ as $F_\beta(\cdot)$.

All signals are discrete-time memoryless sequences of samples drawn
from the same stationary probability distribution. This differs from
the detection scenario often considered in SR research, in which the
input signal is periodic. Such a signal does not convey new
information with an increasing number of samples, and cannot be
considered from an information theoretic
viewpoint~\cite{DeWeese.95}.

Consider two continuous random variables, $X$ and $Y$, with PDFs
$f_X(x)$ and $f_Y(x)$, with the same support, $S$. The {\em relative
entropy}---or Kullback-Liebler divergence---between the two
distributions is defined as~\cite{Cover}
\begin{equation}\label{2:relent_def}
    D(f_X||f_Y) = \int_{\eta\in S} f_X(\eta)\log_2{\left(\frac{f_X(\eta)}{f_Y(\eta)}\right)}d\eta.
\end{equation}
Suppose $X$ and $Y$ have joint PDF, $f_{XY}(x,y)$. Shannon's mutual
information between $X$ and $Y$ is defined as the relative entropy
between the joint PDF and the product of the marginal
PDFs~\cite{Cover},
\begin{align}\label{2:relent_def1}
    I(X,Y) &=
    \int_x\int_yf_{XY}(x,y)\log_2{\left(\frac{f_{XY}(x,y)}{f_X(x)f_Y(y)}\right)}dxdy\notag\\
    &= H(Y)-H(Y|X)\quad\mbox{bits per sample}.
\end{align}
where $H(Y)$ is the entropy of $Y$ and $H(Y|X)$ is the average
conditional entropy of $Y$ given $X$.

The definition of mutual information also holds for discrete random
variables, and for one variable discrete and one continuous. The
entropy of a discrete random variable, $Y$, is given by
\begin{equation}\label{Hz}
    H(Y) =-\sum_{n=0}^NP_Y(n)\log_{2}P_Y(n),
\end{equation}
while a continuous random variable, $X$, has {\em differential}
entropy
\begin{equation}\label{H_cont}
    H(X) = -\int_{\eta\in S} f_X(\eta)\log_2{(f_X(\eta))}d\eta.
\end{equation}
In this paper we are interested in the case of $X$ continuous with
support $S$ and $Y$ discrete, with $N$ states, in which case the
average conditional entropy of $Y$ given $X$ is
\begin{equation}\label{H_YgivenX}
    H(Y|X) = -\int_{x\in
    S}f_X(x)\sum_{n=0}^NP_{Y|X}(n|x)\log_{2}{(P_{Y|X}(n|x))}dx.
\end{equation}

In information theory, the term {\em channel capacity} is defined as
being the maximum achievable mutual information of a given
channel~\cite{Cover}. Suppose $X$ is the source random variable, and
$Y$ is the random variable at the output of the channel. Usually,
the channel is assumed to be fixed and the maximization performed
over all possible source PDFs, $f_X(x)$. The channel capacity, $C$,
can be expressed as the optimization problem,
\begin{equation}\label{4:ChannelCapacityDef}
    \mbox{Find:} \quad C = \max_{\{f_X(x)\}} I(X,Y).
\end{equation}
Usually there are prescribed constraints on the source distribution
such as a fixed average power, or a finite alphabet~\cite{Cover}. In
Sec.~\ref{c:compare} we will also consider the more stringent
constraint that the PDF of the source is known other than its
variance.  In this situation, channel capacity is determined by
finding the optimal source variance, or as is often carried out in
SR research, the optimal noise variance.

\subsection{SSR Model}\label{c:model}

Fig.~\ref{f:4:SSR_Array} shows a schematic diagram of the SSR model.
The array consists of $N$ parallel threshold nonlinearities---or
`devices', each of which receive the same random input signal, $X$,
with PDF $f_X(\cdot)$. The $i$--th device in the model is subject to
continuously valued \textit{iid}---independent and identically
distributed---additive random noise, $\eta_{i}$ ($ i = 1,..,N$),
with PDF $f_\eta(\cdot)$. Each noise signal is required to also be
independent of the signal, $X$. The output of each device, $y_i$, is
unity if the input signal, $X$, plus the noise on that device's
threshold, $\eta_i$, is greater than the threshold value, $\theta$.
The output signal is zero otherwise. The outputs from each device,
$y_i$, are summed to give the overall output signal, $y=
\sum_{i=1}^{N}y_i$. This output is integer valued, $y\in[0,..,N]$,
and is therefore a quantization (digitization) of
$X$~\cite{McDonnell.05}.

The conditional PMF of the output given the input is
$P_{y|X}(y=n|X=x), n\in[0..,N]$. We abbreviate this to
$P_{y|X}(n|x)$. The output distribution is
\begin{equation}\label{4:Py}
  P_y(n) = \int_x P_{y|X}(n{\mid}x)f_X(x)dx \quad n \in
  {0,..,N}.
\end{equation}

The mutual information between $X$ and $y$ is that of a
semi-continuous channel~\cite{Stocks.Mar2000}, and can be written as
\begin{align}\label{5:Info}
  I(X,y) =&~ H(y) - H(y{\mid}X)\notag\\
  =&~ -\sum_{n=0}^N P_y(n)\log_{2}P_y(n)
  -\notag\\
  &\left(-\int_{-\infty}^\infty f_X(x)\sum_{n=0}^N
  P_{y|X}(n{\mid}x)\log_{2}P_{y|X}(n{\mid}x)dx\right).
\end{align}

To progress further we use the notation introduced
in~\cite{Stocks.Mar2000}. Let $P_{1{\mid}x}$ be the probability of
the $i$--th threshold device giving output $y_i=1$ in response to
input signal value, $X=x$. If the noise CDF is $F_\eta(\cdot)$, then
\begin{equation}\label{4:Prob_on1}
    P_{1{\mid}x} = 1-F_\eta(\theta-x).
\end{equation}
As noted in~\cite{Stocks.Mar2000}, $P_{y|X}(n|x)$ is given by the
binomial distribution as
\begin{equation}\label{4:binomial}
P_{y|X}(n|x) = {N\choose n}P_{1|x}^n(1-P_{1|x})^{N-n} \quad n \in
  {0,..,N},
\end{equation}
and Eq.~(\ref{5:Info}) reduces to
\begin{align}\label{5:Info1a}
    I(X,y) =&
    -\sum_{n=0}^NP_y(n)\log_{2}{\left(\frac{P_y(n)}{{N\choose n}}\right)}+\notag\\
    &N\int_xf_X(x)P_{1{\mid}x}\log_2{P_{1{\mid}x}}dx+\notag\\
    &N\int_xf_X(x)(1-P_{1{\mid}x})\log_2{(1-P_{1{\mid}x})}dx.
\end{align}
Numerically evaluating Eq.~(\ref{5:Info1a}) as a function of input
SNR for given signal and noise distributions finds that the mutual
information has a unimodal stochastic resonance curve for $N>1$,
even when the signal and noise are both suprathreshold---i.e. the
threshold value, $\theta$, is set to the signal
mean~\cite{Stocks.01b,McDonnell.02MEJ}.

Further analytical simplification of Eq.~(\ref{5:Info}) is possible
in the case where the signal and noise PDFs are identical with the
same variance, i.e. $f_X(x)=f_\eta(\theta-x)$
$\forall~x$~\cite{Stocks.01b}. The result is
\begin{equation}\label{4:Info2}
    I(X,y)=\log_2{(N+1)}-\frac{N}{2\ln{2}}-\frac{1}{N+1}\sum_{n=2}^N(N+1-2n)\log_2{n}.
\end{equation}
What is quite remarkable about this result is that the mutual
information is independent of the shape of the PDFs of the signal
and noise, other than that $f_X(x)=f_\eta(\theta-x)~\forall~x$. This
means that both PDFs have the same shape, but may possibly have
different means, and be mutually reversed along the $x$-axis about
their means. In Sec.~\ref{c:cons} we compare the mutual information
of Eq.~(\ref{4:Info2}) with our calculations of the general channel
capacity.

\subsection{Describing SSR Using a Single PDF, $f_Q(\tau)$}\label{c:Q}

We now show that the mutual information in the SSR model depends
solely on $N$, and an auxiliary PDF, $f_Q(\cdot)$. This PDF is shown
to be that of the random variable describing the conditional {\em
average} output of the SSR model, given that the input signal is
$X=x$.

\subsubsection{$f_Q(\tau)$ as the PDF of the {\em Average} Transfer
Function}

Although the output of the SSR model, $y$, is a discrete random
variable, the conditional {\em expected} value of $y$, given the
input is $X=x$, {\em is} a continuous random variable, since $X$ is.
We label this random variable as $\bar{Y}$. Since the PMF of $y$
given $X=x$ is the binomial PMF as in Eq.~(\ref{4:binomial}), we
know that $\bar{Y}$ is the random variable that results from
$\bar{y}=\mbox{E}[y|X=x] = NP_{1|x}$. Inverting this gives
$x=\theta-F_\eta^{-1}\left(1-\frac{\bar{y}}{N}\right)$.

The PDF of $\bar{Y}$ can be derived from $f_X(\cdot)$, since
$\bar{y}=NP_{1|x}$ provides an invertible transformation of $X$,
with PDF $f_X(x)$, to $\bar{Y}$, with PDF $f_{\bar{Y}}(\bar{y})$.
Using the well known expression for the resultant PDF, and provided
the support of $f_X(x)$ is contained in the support of
$f_\eta(\theta-x)$---since otherwise $\frac{dx}{d\bar{y}}$ does not
necessarily exist---we have
\begin{align}\label{5:Pyn3}
    f_{\bar{Y}}(\bar{y}) =&
    f_X\left(x\right)\left|\frac{dx}{d\bar{y}}\right|\Big |_{x=\theta-F_\eta^{-1}\left(1-\frac{\bar{y}}{N}\right)}\notag\\
    =&\frac{f_X(x)}{Nf_\eta(\theta-x)}\Big |_{x=\theta-F_\eta^{-1}\left(1-\frac{\bar{y}}{N}\right)},~\quad \bar{y}\in[0,N].
\end{align}
Our condition regarding the supports of the signal and noise ensures
that $f_\eta(\cdot)\ne 0$. If we make a further change to a new
random variable, $Q$, via $\tau=\frac{\bar{y}}{N}$, the PDF of $Q$
is
\begin{equation}\label{4:Q_def}
    f_Q(\tau) = \frac{f_X(x)}{f_\eta(\theta-x)}\Big
    |_{x=\theta-F_\eta^{-1}(1-\tau)},\quad\tau\in[0,1],
\end{equation}
and the PDF of $\bar{Y}$ can be written as
\begin{equation}\label{5:Pyn_Q0}
    f_{\bar{Y}}(\bar{y}) =
    \frac{f_Q\left(\frac{\bar{y}}{N}\right)}{N},
\end{equation}
which illustrates the physical significance of the auxiliary PDF,
$f_Q(\cdot)$, as the PDF of $\frac{\bar{y}}{N}$.

\subsubsection{Mutual Information in Terms of $f_Q(\tau)$}

Making a change of variable in Eq.~(\ref{5:Info1a}) from $x$ to
$\tau$, via $\tau=P_{1|x}=1-F_\eta(\theta-x)$ gives
\begin{align}\label{4:pstar_n1}
    I(X,y)=&-\sum_{n=0}^N
    P_y(n)\log_{2}{\left(\frac{P_y(n)}{{N\choose n}}\right)}+\notag\\
    &N\int_{\tau=0}^{\tau=1}f_Q(\tau)\tau\log_2{\tau}d\tau+\notag\\
    &N\int_{\tau=0}^{\tau=1}f_Q(\tau)(1-\tau)\log_2{(1-\tau)}d\tau,
\end{align}
where
\begin{equation}\label{4:pstar_n}
    P_y(n)={N\choose n}\int_{\tau=0}^{\tau=1}f_Q(\tau)\tau^n(1-\tau)^{N-n}d\tau.
\end{equation}
Eqs.~(\ref{4:pstar_n1}) and~(\ref{4:pstar_n}) show that the PDF
$f_Q(\tau)$ encapsulates the behavior of the mutual information in
the SSR model.

\subsubsection{Entropy of the random variable, $Q$}

If we make a change of variable from $\tau$ to $x$, and note that
$f_X(x)dx=f_Q(\tau)d\tau$, the entropy of $Q$ can be written as
\begin{align}\label{H_Q_def}
    H(Q)&=-\int_0^1 f_Q(\tau)\log_2{(f_Q(\tau))}d\tau\notag\\
    &=-\int_xf_X(x)\log_2{\left(\frac{f_X(x)}{f_\eta(\theta-x)}\right)}dx\notag\\
    &=-D(f_X(x)||f_\eta(\theta-x)),
\end{align}
which is the negative of the relative entropy between the signal
PDF, and the noise PDF reversed about $x=0$ and shifted by $\theta$.
In the event that the noise PDF is an even function about its mean,
and $\theta$ is equal to the signal mean, then the entropy of $Q$ is
simply the negative of the relative entropy between the signal and
noise PDFs.

\subsubsection{Examples of the PDF $f_Q(\tau)$}

The PDF $f_Q(\tau)$ can be derived for specific signal and noise
distributions. Table~\ref{4:table3} lists $f_Q(\tau)$ for several
cases where the signal and noise share the same distribution and a
mean of zero, but with not necessarily equal variances. The
threshold value, $\theta$, is also set to zero.

For each case considered, the standard deviation of the noise can be
written as  $a\sigma_\eta$, where $a$ is a positive constant, and
the standard deviation of the signal can be written $a\sigma_x$.  We
find that $f_Q(\tau)$ in each case is a function of a single
parameter that we call the {\em noise intensity},
$\sigma=\sigma_\eta/\sigma_x$. Given this, from
Eq.~(\ref{4:pstar_n1}), it is clear that the mutual information must
be a function only of the ratio, $\sigma$, so that it is invariant
to a change in $\sigma_x$ provided $\sigma_\eta$ changes by the same
proportion. This fact is noted to be true for the Gaussian case
in~\cite{Stocks.Mar2000}, and the uniform case in~\cite{Stocks.01b},
but here we have illustrated why.

We note however, that if $\theta$ is not equal to the signal mean,
then $f_Q(\tau)$ will depend on the ratio $\frac{\theta}{\sigma_x}$,
as well as $\theta$ and $\sigma$, and therefore so will the mutual
information.

Table~\ref{4:table3} also lists the entropy of $Q$ for three cases
where an analytical expression could be found.

\subsection{Large $N$ SSR: Literature Review and Outline of This Paper}\label{c:5litrev}

In the absence of noise, the maximum mutual information is the
maximum entropy of the output signal, $\log_2{(N+1)}$. It has been
shown for very specific signal and noise distributions that the
mutual information in the SSR model scales with $0.5\log_2{(N)}$ for
large $N$~\cite{Stocks.01a,Stocks.01b}. This means that the channel
capacity for large $N$ under the  specified conditions is about half
the maximum noiseless channel capacity. This situation is discussed
in Sec.~\ref{c:compare}.

The only other work to consider SSR in the large $N$ regime finds
that the optimal noise intensity for Gaussian signal and noise
occurs for $\sigma\simeq0.6$~\cite{Hoch.03a}.
Unlike~\cite{Stocks.01a}---which uses the exact expression of
Eq.~(\ref{4:Info2}), and derives a large $N$ expression by
approximating the summation with an integral---\cite{Hoch.03a}
begins by using a {\it Fisher information} based approximation to
the mutual information.

In Appendix~\ref{c:5Fisher} we re-derive the formula
of~\cite{Hoch.03a} in a different manner, which results in new large
$N$ approximations for the {\it output entropy}, as well as the
mutual information. These approximations provide the basis for the
central result of this paper, which is a general sufficient
condition for achieving channel capacity in the SSR model, for any
arbitrary specified signal or noise distribution. This is discussed
in Section~\ref{c:5capacity}. These new general results are compared
with the specific results of~\cite{Stocks.01a,Stocks.01b,Hoch.03a}
in Sec.~\ref{c:compare}.

\section{A General Expression for the SSR Channel Capacity for Large $N$}\label{c:5capacity}

Fisher information~\cite{Lehmann,Cover} has previously been
discussed in numerous papers on both neural coding~\cite{Bethge.02}
and stochastic resonance~\cite{Greenwood.04}, and
both~\cite{Stemmler.96,Greenwood.00}. However, most SR studies using
Fisher information consider only the case where the signal itself is
not a random variable. When it is a random variable, it is possible
to connect Fisher information and Shannon mutual information under
special conditions, as discussed
in~\cite{Stemmler.96,Brunel.98,Kang.01,Hoch.03a}.

It is demonstrated in~\cite{Hoch.03a} that the Fisher information at
the output of the SSR model as a function of input signal value
$X=x$, is given by
\begin{equation}\label{5:Fisher_I}
    J(x) = \left(\frac{dP_{1|x}}{dx}\right)^2\frac{N}{P_{1|x}(1-P_{1|x})}.
\end{equation}
In~\cite{Hoch.03a}, Eq.~(\ref{5:Fisher_I}) is used to approximate
the large $N$ mutual information in the SSR model via the formula
\begin{align}\label{5:Info_largeN1}
    I(X,y)
    &=H(X)-0.5\int_{x=-\infty}^{x=\infty}f_X(x)\log_2{\left(\frac{2{\pi}e}{J(x)}\right)}dx.
\end{align}
This expression---which is derived under much more general
circumstance in~\cite{Clarke.90,Brunel.98}---relies on an assumption
that an efficient Gaussian estimator for $x$ can be found from the
output of the channel, in the limit of large $N$.

In Appendix~\ref{c:5Fisher} we outline an alternative derivation to
Eq.~(\ref{5:Info_largeN1})---from which Eq.~(\ref{5:Fisher_I}) can
be inferred---that is specific to the SSR model, and provides
additional justification for its large $N$ asymptotic validity. This
alternative derivation allows us to find individual expressions for
both the output entropy and conditional output entropy. This
derivation makes use of the auxiliary PDF, $f_Q(\tau)$, derived in
Sec.~\ref{c:Q}. The significance of this approach is that it leads
to our demonstration of the new results that the output entropy can
be written for large $N$ as
\begin{align}\label{new}
    H(y) &\simeq \log_2{(N)}-D(f_X(x)||f_\eta(\theta-x)),
\end{align}
while the mutual information can be written as
\begin{equation}\label{DP_S}
    I(X,y)
    \simeq0.5\log_2{\left(\frac{N\pi}{2e}\right)}-D(f_X||f_S),
\end{equation}
where $f_S(\cdot)$ is a PDF known as Jeffrey's prior,
\begin{equation}\label{5:sqrtJ2}
    f_S(x) = \frac{\sqrt{J(x)}}{\pi\sqrt{N}}.
\end{equation}
It is proven in Appendix~\ref{c:A5Jeffrey} that for the SSR model
Eq.~(\ref{5:sqrtJ2}) is indeed a PDF. This is a remarkable result,
as in general Jeffrey's prior has no such simple form. Substitution
of Eq.~(\ref{5:sqrtJ2}) into Eq.~(\ref{DP_S}) and simplifying leads
to Eq.~(\ref{5:Info_largeN1}), which verifies this result.

By inspection of Eq.~(\ref{5:Fisher_I}), $f_S(x)$ can be derived
from knowledge of the noise PDF, $f_\eta(\eta)$, since
\begin{equation}\label{f_S}
    f_S(x) = \frac{f_\eta(\theta-x)}{\pi\sqrt{F_\eta(\theta-x)(1-F_\eta(\theta-x))}}.
\end{equation}

\subsection{A Sufficient Condition for Optimality}

Since relative entropy is always non-negative, from Eq.~(\ref{DP_S})
a sufficient condition for achieving the large $N$ channel capacity
is that
\begin{equation}\label{sufficient}
    f_X(x)=f_S(x)\quad\forall~x,
\end{equation}
with the resultant capacity as
\begin{equation}\label{5:sqrtJ5}
    C(X,y) = 0.5\log_2{\left(\frac{N\pi}{2e}\right)}\simeq 0.5\log_2{N}-0.3956.
\end{equation}
Eq.~(\ref{5:sqrtJ5}) holds provided the conditions for the
approximation given by Eq.~(\ref{5:Info_largeN1}) hold. Otherwise,
the RHSs of Eqs.~(\ref{new}) and~(\ref{DP_S}) give lower bounds.
This means that for the situations considered previously
in~\cite{Hoch.03a,Stocks.01a} where the signal and noise both have
the same distribution (but different variances), we can expect to
find channel capacity that is less than or equal to that of
Eq.~(\ref{5:sqrtJ5}). This is discussed in Sec.~\ref{c:compare}.

The derived sufficient condition of Eq.~(\ref{sufficient}) leads to
two ways in which capacity can be achieved, (i) an optimal signal
PDF for a given noise PDF, and (ii) an optimal noise PDF for a given
signal PDF.

\subsection{Optimizing the Signal Distribution}\label{c:optsig}

Assuming Eq.~(\ref{5:Info_largeN1}) holds, the channel capacity
achieving input PDF, $f_X^o(x)$, can be found for any given noise
PDF from Eqs.~(\ref{f_S}) and~(\ref{sufficient}) as
\begin{equation}\label{f_X_opt}
    f_X^o(x) = \frac{f_\eta(\theta-x)}{\pi\sqrt{F_\eta(\theta-x)(1-F_\eta(\theta-x))}}.
\end{equation}

\subsubsection{Example: Uniform Noise}

Suppose the {\it iid} noise at the input to each threshold device in
the SSR model is uniformly distributed on the interval
$[-\sigma_\eta/2,\sigma_\eta/2]$ so that it has PDF
\begin{equation}\label{uniform_noise}
    f_\eta(\xi) =
    \frac{1}{\sigma_\eta},\quad~\xi\in[-\sigma_\eta/2,\sigma_\eta/2].
\end{equation}
Substituting Eq.~(\ref{uniform_noise}) and its associated CDF into
Eq.~(\ref{f_X_opt}), we find that the optimal signal PDF is
\begin{equation}\label{opt_sig}
    f_X^o(x)= \frac{1}{\pi\sqrt{\frac{\sigma_\eta^2}{4}-(x-\theta)^2}},\quad
    x\in[\theta-\sigma_\eta/2,\theta+\sigma_\eta/2].
\end{equation}
This PDF is in fact the PDF of a sine-wave with uniformly random
phase, amplitude $\sigma_\eta/2$, and mean $\theta$. A change of
variable to the interval $\tau\in[0,1]$ via the substitution
$\tau=(x-\theta)/\sigma_\eta+0.5$ results in the PDF of the Beta
distribution with parameters $0.5$ and $0.5$, also known as the
arcsine distribution. As mentioned in Sec.~\ref{s:intro}, this
result provides some theoretical justification for the
analog-to-digital conversion method proposed in~\cite{Nguyen.06}.

This Beta distribution is bimodal, with the most probable values of
the signal those near zero and unity. Similar results for an optimal
input distribution in an information theoretic optimization of a
neural system have been found in~\cite{Schreiber.02}. These results
were achieved numerically using the Blahut-Arimoto algorithm often
used in information theory to find channel capacity achieving source
distributions, or rate-distortion functions~\cite{Cover}.

\subsubsection{Gaussian Noise}

Suppose the {\it iid} noise at the input to each threshold device
has a zero mean Gaussian distribution with variance $\sigma_\eta^2$,
with PDF
\begin{equation}\label{Gaussian_noise}
    f_\eta(\xi) =
    \frac{1}{\sqrt{2\pi\sigma_\eta^2}}\exp{\left(-\frac{\xi^2}{2\sigma_\eta^2}\right)}.
\end{equation}
Substituting Eq.~(\ref{Gaussian_noise}) and its associated CDF into
Eq.~(\ref{f_X_opt}), gives the optimal signal PDF.  The resultant
expression for $f_X^o(x)$ does not simplify much, and contains the
standard error function, $\mbox{erf}(\cdot)$~\cite{Spiegel}.

We are able to verify that the resultant PDF has the correct shape
via Fig.~8 in~\cite{Hoch.03a}, which presents the result of {\em
numerically} optimizing the signal PDF, $f_X(x)$, for unity variance
zero mean Gaussian noise, $\theta=0$, and $N=10000$. As with the
work in~\cite{Schreiber.02}, the numerical optimization is achieved
using the Blahut-Arimoto algorithm.   It is remarked
in~\cite{Hoch.03a} that the optimal $f_X(x)$ is close to being
Gaussian. This is illustrated by plotting both $f_X(x)$ and a
Gaussian PDF with nearly the same peak value as $f_X(x)$. It is
straightforward to show that a Gaussian with the same peak value as
our analytical $f_X^o(x)$ has variance $0.25\pi^2$.  If the signal
was indeed Gaussian, then we would have $\sigma=2/\pi\simeq 0.6366$,
which is very close to the value calculated for actual Gaussian
signal and noise in Sec.~\ref{c:compare}.

\begin{figure}[htbp]
\centering\includegraphics[scale=0.5]{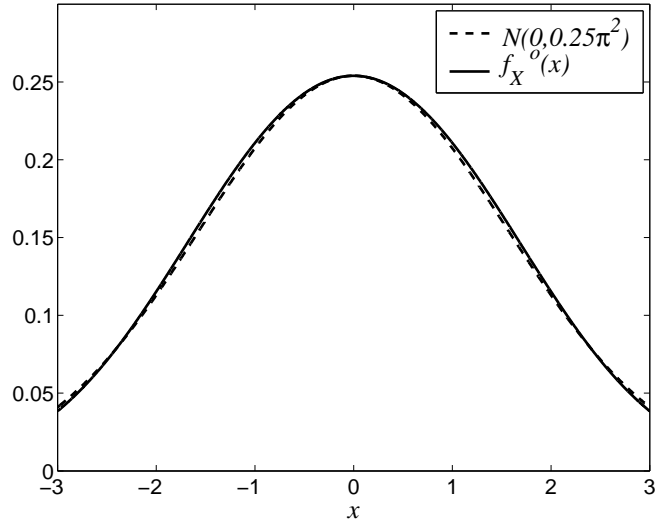} \caption[]{The
optimal signal PDF, $f_X^o(x)$, for zero mean, unity variance
Gaussian noise, and threshold value $\theta=0$, as obtained from
Eq.~(\ref{f_X_opt}). Superimposed is a Gaussian PDF with the same
peak value as $f_X^o(x)$, so that it has variance $0.25\pi^2$. This
figure uses our new theoretical results to analytically replicate
Fig.~8 in~\cite{Hoch.03a}, which was calculated numerically.}
\label{f:Opt_f_X}
\end{figure}

Our analytical $f_X^o(x)$ from Eqs.~(\ref{Gaussian_noise})
and~(\ref{f_X_opt}), with $\theta=0$, is plotted on the interval
$x\in[-3,3]$ in Fig.~\ref{f:Opt_f_X}, along with a Gaussian PDF with
variance $0.25\pi^2$. Clearly the optimal signal PDF is very close
to the Gaussian PDF. Our Fig.~\ref{f:Opt_f_X} is virtually identical
to Fig.~8 in~\cite{Hoch.03a}. It is emphasized that the results
in~\cite{Hoch.03a} were obtained using an entirely different method
that involves numerical iterations, and therefore provides excellent
validation of our theoretical results.

\subsection{Optimizing the Noise Distribution}

We now assume that the signal distribution is known and fixed. We
wish to achieve channel capacity by finding the optimal noise
distribution.  It is easy to show by integrating Eq.~(\ref{f_S})
that the CDF corresponding to the PDF, $f_S(\cdot)$, evaluated at
$x$, can be written in terms of the CDF of the noise distribution as
\begin{equation}\label{F_s}
F_S(x) =
1-\frac{2}{\pi}\arcsin{\left(\sqrt{F_\eta(\theta-x)}\right)}.
\end{equation}
If we now let $f_X(x)=f_S(x)$, then $F_X(s)=F_S(x)$, and rearranging
Eq.~(\ref{F_s}) gives the optimal noise CDF in terms of the signal
CDF as
\begin{equation}\label{hh}
    F_\eta^o(x) = \sin^2{\left(\frac{\pi}{2}(1-F_X(\theta-x))\right)} = 0.5+0.5\cos{(\pi F_X(\theta-x))}.
\end{equation}
Differentiating $F_\eta^o(x)$ gives the optimal noise PDF as a
function of the signal PDF and CDF,
\begin{equation}\label{hh1}
    f_\eta^o(x) = \frac{\pi}{2}\sin{\left(\pi(1-F_X(\theta-x))\right)}f_X(\theta-x).
\end{equation}
Unlike optimizing the signal distribution, which is the standard way
for achieving channel capacity in information theory~\cite{Cover},
we have assumed a signal distribution, and found the `best' noise
distribution, which is equivalent to optimizing the channel, rather
than the signal.

\subsubsection{Example: Uniform Signal}

Suppose the signal is uniformly distributed on the interval
$x\in[-\sigma_x/2,\sigma_x/2]$. From Eqs.~(\ref{hh})
and~(\ref{hh1}), the capacity achieving noise distribution has CDF
\begin{equation}\label{opt_cdf}
    F_\eta^o(x) = 0.5+0.5\sin{\left(\frac{\pi
    (x-\theta)}{\sigma_x}\right)},\quad~x\in[\theta-\sigma_x/2,\theta+\sigma_x/2]
\end{equation}
and PDF
\begin{equation}\label{opt_pdf}
    f_\eta^o(x) = \frac{\pi}{2\sigma_x}\cos{\left(\frac{\pi
    (x-\theta)}{\sigma_x}\right)},\quad~x\in[\theta-\sigma_x/2,\theta+\sigma_x/2].
\end{equation}
Substitution of $F_\eta^o(x)$ and $f_\eta^o(x)$ into
Eq.~(\ref{5:Fisher_I}) finds the interesting result that the Fisher
information is constant for all $x$,
\begin{equation}\label{opt_J}
    J(x) = N\frac{\pi^2}{\sigma_x^2}.
\end{equation}
This is verified in Eq.~(\ref{fisher_P}) below.

\subsection{Consequences of Optimizing the Large $N$ Channel
Capacity}\label{c:cons}

\subsubsection{Optimal Fisher Information}

Regardless of whether we optimize the signal for given noise, or
optimize the noise for a given signal, it is straightforward to show
that the Fisher information can be written as a function  of the
signal PDF,
\begin{equation}\label{fisher_P}
    J(x) = N\pi^2(f_X(x))^2.
\end{equation}
Therefore, the Fisher information at large $N$ channel capacity is
constant for the support of the signal iff the signal is uniformly
distributed.  The optimality of constant Fisher information in a
neural coding context is studied in~\cite{Bethge.02}.

\subsubsection{The Optimal PDF $f_Q(\tau)$}

A further consequence that holds in both cases is that the ratio of
the signal PDF to the noise PDF is
\begin{equation}\label{ratio}
    \frac{f_X(x)}{f_\eta(\theta-x)} = \frac{2}{\pi\sin{(\pi (1-F_X(x)))}}.
\end{equation}
This is not a PDF. However, if we make a change of variable via
$\tau=1-F_\eta(\theta-x)$ we get the PDF $f_Q(\tau)$ discussed in
Sec.~\ref{c:Q}, which for channel capacity is
\begin{equation}\label{OptQ}
    f_Q^o(\tau) = \frac{1}{\pi\sqrt{\tau(1-\tau)}},\quad~\tau\in[0,1].
\end{equation}
This optimal $f_Q(\tau)$ is in fact the PDF of the beta distribution
with parameters $0.5$ and $0.5$, i.e. the arcsine distribution.  It
is emphasised that this result holds {\em regardless} of whether the
signal PDF is optimised for a given noise PDF or {\it vice versa}.

\subsubsection{Output Entropy at Channel Capacity}

From Eq.~(\ref{H_Q_def}), the entropy of $Q$ is equal to the
negative of the relative entropy between $f_X(x)$ and
$f_\eta(\theta-x)$. The entropy of $Q$ when capacity is achieved can
be calculated from Eq.~(\ref{OptQ}) using direct integration as
\begin{equation}\label{H_Q}
    H(Q) = \log_2{(\pi)}-2.
\end{equation}
From Eqs.~(\ref{new}) and~(\ref{H_Q_def}), the large $N$ output
entropy at channel capacity in the SSR model is
\begin{equation}\label{H_Q1}
H(y)=\log_2{\left(\frac{N\pi}{4}\right)}.
\end{equation}

\subsubsection{The Optimal Output PMF is Beta-Binomial}

Suppose we have signal and noise such that
$f_Q(\tau)=f_Q^o(\tau)$---i.e. the signal and noise satisfy the
sufficient condition, Eq.~(\ref{sufficient})---but that $N$ is not
necessarily large. We can derive the output PMF for this situation,
by substituting Eq.~(\ref{OptQ}) into Eq.~(\ref{4:pstar_n}) to get
\begin{align}\label{beta_bino}
    P_y(n)
    &=\left(^N_n\right)\frac{1}{\pi}\int_0^1\tau^{(n-0.5)}(1-\tau)^{(N-n-0.5)}d\tau\notag\\
    &=\left(^N_n\right)\frac{\beta(n+0.5,N-n+0.5)}{\beta(0.5,0.5)}.
\end{align}
where $\beta(a,b)$ is a Beta function.  This PMF can be recognized
as that of the Beta-binomial---or negative
hypergeometric---distribution with parameters $N,0.5$,
$0.5$~\cite{Evans}. It is emphasized that Eq.~(\ref{beta_bino})
holds as an exact analytical result for {\em any} $N$.

\subsubsection{Analytical Expression for the Mutual Information}

\begin{figure}[htbp]
\begin{center}
{\subfigure[~Exact
Expressions]{\includegraphics[scale=0.4]{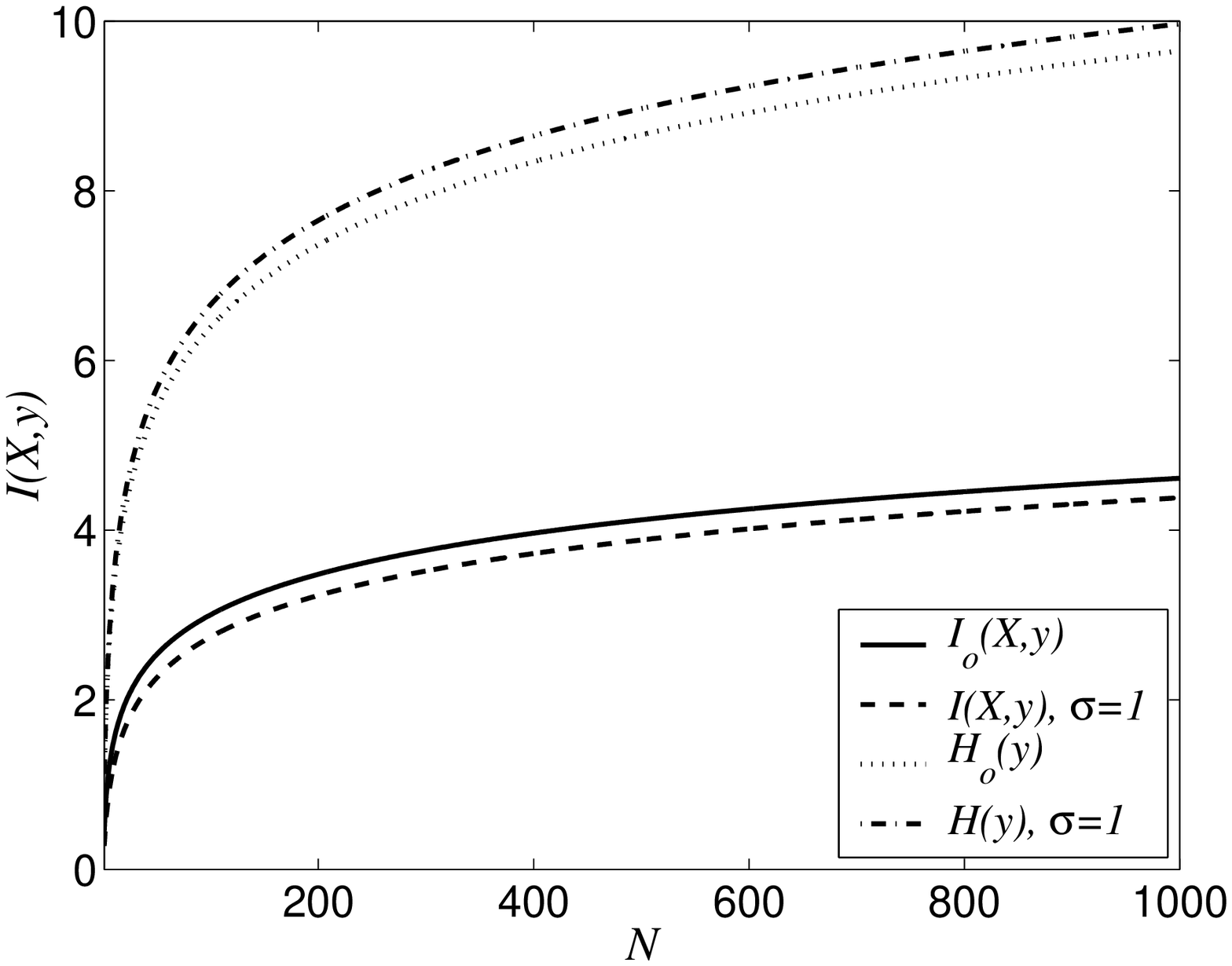}\label{f:Capacity1}}
~ \subfigure[~Error in large $N$
expressions]{\includegraphics[scale=0.4]{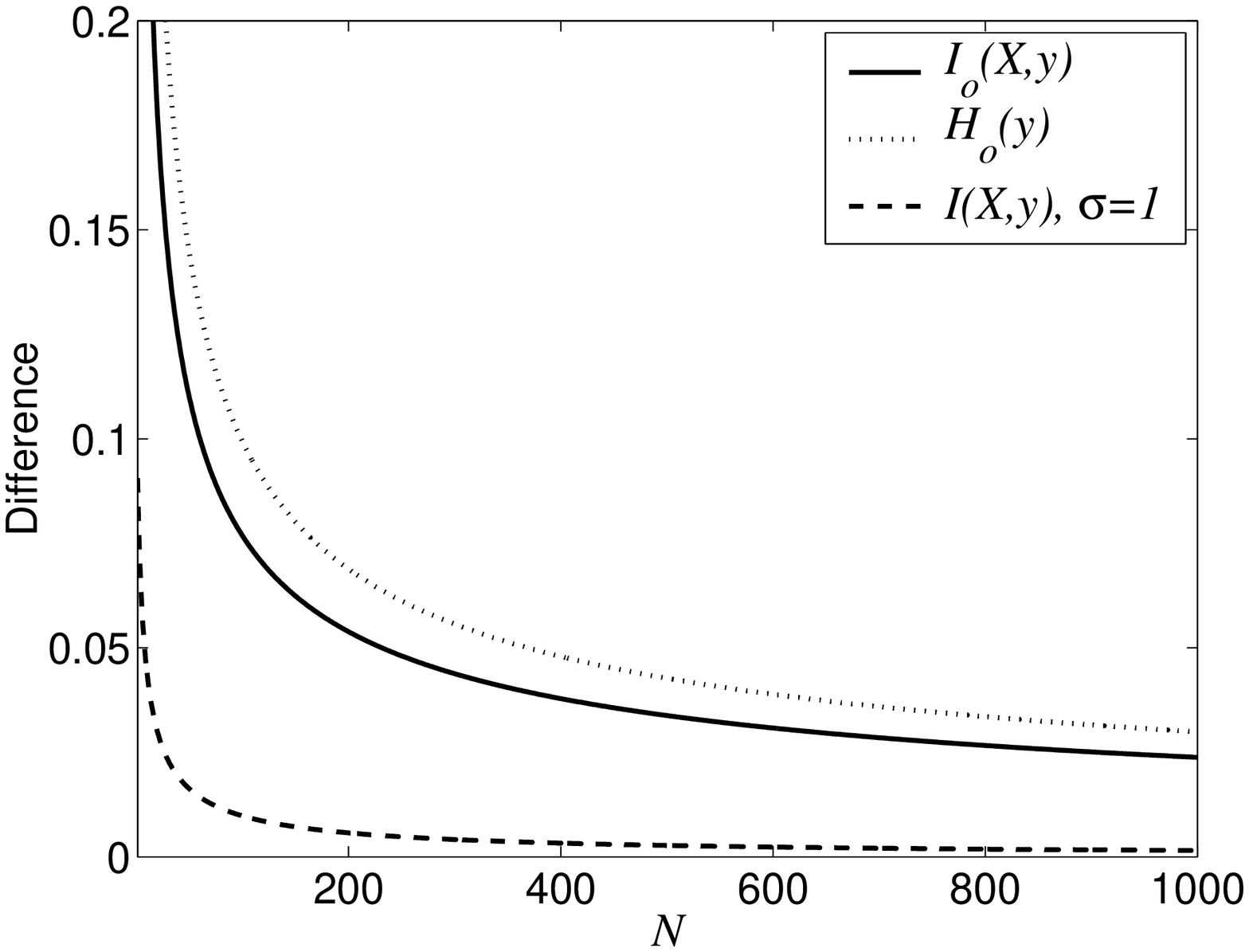}\label{f:Capacity2}}
} \caption[]{(a) Exact expressions obtained using $f_Q^o(\tau)$, for
$I_o(X,y)$, and $H_o(y)$, as well as the exact mutual information
and output entropy when $f_X(x)=f_\eta(\theta-x)$ (denoted as
$\sigma=1$), as a function of $N$. (b) The difference between the
exact expressions for $I_o(X,y)$, $H_o(y)$ and $I(X,y)$ for
$f_X(x)=f_\eta(\theta-x)$, and the corresponding large $N$
expressions given by Eqs.~(\ref{DP_S}),~(\ref{H_Q1})
and~(\ref{5:InfoNew1}).}\label{f:Capacity}
\end{center}
\end{figure}

The exact expression for the output PMF of Eq.~(\ref{beta_bino})
allows exact calculation of both the output entropy, and the mutual
information without need for numerical integration, using
Eq.~(\ref{4:pstar_n1}). This is because when
$f_Q(\tau)=f_Q^o(\tau)$, the integrals in Eq.~(\ref{4:pstar_n1}) can
be evaluated exactly to get
\begin{align}\label{Info_betabin}
    I_o(X,y)=&-\sum_{n=0}^N
    P_y(n)\log_{2}{\left(\frac{P_y(n)}{{N\choose n}}\right)}+N\log_2{\left(\frac{e}{4}\right)}.
\end{align}
The exact values of $I_o(X,y)$ and the corresponding output entropy,
$H_o(y)$, are plotted in Fig.~\ref{f:Capacity1} for $N=1,..,1000$.
For comparison, the exact $I(X,y)$ of Eq.~(\ref{4:Info2}), which
holds for $f_X(x)=f_\eta(\theta-x)$, is also plotted, as well as the
corresponding entropy, $H(y)=\log_2{(N+1)}$. It is clear that
$I_o(X,y)$ is always larger than the mutual information of the
$f_X(x)=f_\eta(\theta-x)$ case, and that $H_o(y)$ is always less
than its entropy, which is the maximum output entropy.

To illustrate that the large $N$ expressions derived are lower
bounds to the exact formula plotted in Fig.~\ref{f:Capacity1}, and
that the error between them decreases with $N$,
Fig.~\ref{f:Capacity2} shows the difference between the exact and
the large $N$ mutual information and output entropy. This difference
clearly decreases with increasing $N$.

\subsection{A Note on the Output Entropy}

The SSR model has been described in terms of signal quantization
theory in~\cite{McDonnell.05}, and compared with the related process
of companding in~\cite{Amblard.06}. In this context {\em
quantization} means the conversion of a continuously valued signal
to a discretely valued signal that has only a finite number of
possible values. Quantization in this sense occurs in
analog-to-digital converter circuits, lossy compression algorithms,
and in histogram formation~\cite{Gersho}. For a deterministic scalar
quantizer with $N+1$ output states, $N$ threshold values are
required. In quantization theory, there is a concept of high
resolution quantizers, in which the distribution of
$N\rightarrow\infty$ threshold values can be described by a point
density function, $\lambda(x)$. For such quantizers, it can be shown
that the quantizer output, $y$, in response to a random variable,
$X$, has entropy $H(y) \simeq
\log_2{N}-D(f_X||\lambda)$~\cite{Gersho}. This is strikingly similar
to our Eq.~(\ref{new}) for the large $N$ output entropy of the SSR
model. In fact, since the noise that perturbs the fixed threshold
value, $\theta$, is additive, each threshold acts as an {\it iid}
random variable with PDF $f_\eta(\theta-x)$, and therefore for large
$N$, $f_\eta(\theta-x)$ acts as a density function describing the
relative frequency of threshold values as a function of $x$, just as
$\lambda(x)$ does for a high resolution deterministic quantizer.

For deterministic quantizers, the point density function can be used
to approximate the high resolution distortion incurred by the
quantization process. For the SSR model however, since the
quantization has a random aspect, the distortion has a component due
to randomness as well as lossy compression, and cannot be simply
calculated from $f_\eta(\cdot)$. Instead, one can use the Fisher
information to calculate the asymptotic mean square error
distortion, which is not possible for deterministic high resolution
quantizers.

\section{Channel Capacity for Large $N$ and `Matched' Signal and
Noise}\label{c:compare}

Unlike the previous section, we now consider channel capacity under
the constraint of `matched' signal and noise distributions---i.e.
where both the signal and noise, while still independent, have the
same distribution, other than their variances. The mean of both
signal and noise is zero and the threshold value is also $\theta=0$.
In this situation the mutual information depends solely on the ratio
$\sigma = \sigma_\eta/\sigma_x$, which is the only free variable.
Finding channel capacity is therefore equivalent to finding the
optimal value of noise intensity, $\sigma$. Such an analysis
provides verification of the more general capacity expression of
Eq.~(\ref{5:sqrtJ5}), which cannot be exceeded.

Furthermore, inspection of Eq.~(\ref{5:Info_largeN_3}) shows that
the large $N$ approximation to the mutual information consists of a
term that depends on $N$ and a term that depends only on $\sigma$.
This shows that for large $N$ the channel capacity occurs for the
same value of $\sigma$---which we denote as $\sigma_o$---for all
$N$.

This fact is recognized in both~\cite{Stocks.01a} for uniform signal
and noise---where $\sigma_o\rightarrow 1$---and~\cite{Hoch.03a}, for
Gaussian signal and noise. Here, we investigate the value of
$\sigma_o$ and the mutual information at $\sigma_o$ for other signal
and noise distributions, and compare the channel capacity obtained
with the case where $f_X(x)=f_S(x)$. This comparison finds that the
results of~\cite{Hoch.03a} overstates the true capacity, and that
large $N$ results in~\cite{Stocks.01a,Stocks.01b} need to be
improved to be consistent with the central results of this paper.

From Eq.~(\ref{DP_S}), channel capacity for large $N$ occurs for the
value of $\sigma$ that minimizes the relative entropy between $f_X$
and $f_S$. If we let
\begin{equation}\label{blah}
    f(\sigma)=
    \int_{x=-\infty}^{x=\infty}f_X(x)\ln{\left(\frac{1}{J(x)}\right)}dx,
\end{equation}
then from Eq.~(\ref{5:Info_largeN1}), it is also clear that this
minimization is equivalent to solving the following problem,
\begin{equation}\label{5:ChannelCapacity}
     \sigma_o = \min_{\sigma} f(\sigma).
\end{equation}
This is exactly the formulation stated in~\cite{Hoch.03a}.
Problem~(\ref{5:ChannelCapacity}) can be equivalently expressed as
\begin{equation}\label{5:ChannelCapacity2}
     \sigma_o = \min_{\sigma} \left\{f(\sigma)= D(f_X||f_\eta)
     +\int_{x=-\infty}^{x=\infty}f_X(x)\log_2{(P_{1|x})}dx\right\},
\end{equation}
where we have assumed that both the signal and noise PDFs are even
functions. The function $f(\sigma)$ can be found for any specified
signal and noise distribution by numerical integration, and
Problem~(\ref{5:ChannelCapacity2}) easily solved numerically. If an
exact expression for the relative entropy term is known, then only
$g(\sigma)=\int_{x=-\infty}^{x=\infty}f_X(x)\log_2{(P_{1|x})}dx$
needs to be numerically calculated.

Table~\ref{5:table1} gives the result of numerically calculating the
value of $\sigma_o$, and the corresponding large $N$ channel
capacity, $C(X,y)$, for a number of distributions.  In each case,
$C(X,y)-0.5\log_2{(N)} < -0.3956$, as required by
Eq.~(\ref{5:sqrtJ5}). The difference between capacity and
$0.5\log_2{(N)}$ is about $0.4$ bits per sample. In the limit of
large $N$, this shows that capacity is almost identical, regardless
of the distribution. However, the value of $\sigma_o$ at which this
capacity occurs is different in each case.

As discussed in Sec.~\ref{c:model}, the mutual information is
identical whenever the signal and noise PDFs are identical, i.e.
$\sigma=1$. It is shown below in Eq.~(\ref{5:InfoNew1a}) that for
large $N$ the mutual information at $\sigma=1$ is
$I(X,y)=0.5\log_2{(N)}-0.6444$. Given that the channel capacity is
slightly larger than this, as indicated by Table~\ref{5:table1}, for
each case there is a constant difference between the channel
capacity and the mutual information at $\sigma=1$. This value is
also listed in Table~\ref{5:table1}.

\subsection{Improvements to Previous Large $N$ Approximations}

We now use the results of Sec.~\ref{c:5capacity} to show that
previous large $N$ expressions for the mutual information in the
literature for the $\sigma=1$, Gaussian and uniform cases can be
improved.

\subsubsection{SSR for Large $N$ and $\sigma=1$}\label{c:5sig1}

We now consider the situation where $f_X(x)=f_\eta(x)$, so that
$\sigma=1$. It is shown in~\cite{Stocks.01b} that in this case as
$N$ approaches infinity, Eq.~(\ref{4:Info2}) reduces to
\begin{equation}\label{5:Stocks_largeN}
    I(X,y) \simeq 0.5\log_2{\left(\frac{N+1}{e}\right)} \simeq 0.5\log_2{(N+1)}-0.7213.
\end{equation}
To show that this expression can be improved, we begin with the
version of Eq.~(\ref{5:Info_largeN1}) given by
Eq.~(\ref{5:Info_largeN_3}). When $\sigma=1$ we have $f_Q(\tau)=1$
and $H(Q)=0$.  The integrals in Eq.~(\ref{5:Info_largeN_3}) can be
solved to give the large $N$ mutual information at $\sigma=1$ as
\begin{equation}\label{5:InfoNew1a}
    I(X,y)
    \simeq 0.5\log_2{\left(\frac{Ne}{2\pi}\right)} \simeq 0.5\log_2{N}-0.6044.
\end{equation}
Although Eqs.~(\ref{5:Stocks_largeN}) and~(\ref{5:InfoNew1a}) agree
as $N\rightarrow\infty$, the constant terms do not agree. It is
shown in Appendix~\ref{A:Hy_x} that the discrepancy can be resolved
by improving on the approximation to the average conditional
entropy, $H(y|X)$, made in~\cite{Stocks.01b}. The output entropy at
$\sigma=1$ can be shown to be simply
$H(y)=\log_2{(N+1)}$~\cite{Stocks.01b}. Subtracting
Eq.~(\ref{5A:Hy_xNew}) from $H(y)$ and letting $N$ approach infinity
gives
\begin{equation}\label{5:InfoNew1}
    I(X,y)
    \simeq 0.5\log_2{\left(\frac{(N+2)e}{2\pi}\right)},
\end{equation}
which does have a constant term which agrees with
Eq.~(\ref{5:InfoNew1a}). The explanation of the discrepancy is
that~\cite{Stocks.01b} uses the Euler-Maclaurin summation formula to
implicitly calculate $\log_2{(N!)}$ in the large $N$ approximation
to $H(y|X)$. Using Stirling's approximation for $N!$, as done here,
gives a more accurate approximation.

The increased accuracy of Eq.~(\ref{5:InfoNew1a}) can be verified by
numerically comparing both Eq.~(\ref{5:InfoNew1a}) and
Eq.~(\ref{5:Stocks_largeN}) with the exact expression for $I(X,y)$
of Eq.~(\ref{4:Info2}), as $N$ increases. The error between the
exact expression and Eq.~(\ref{5:InfoNew1a}) approaches zero as $N$
increases, whereas the error between Eq.~(\ref{4:Info2}) and
Eq.~(\ref{5:Stocks_largeN}) approaches a nonzero constant for large
$N$ of $0.5\log_2{\left(\frac{e^2}{2\pi}\right)}\simeq 0.117$ bits
per sample.

\subsubsection{Uniform Signal and
Noise}\label{c:5sig1_uniform}

A derivation is given in~\cite{Stocks.01a} of an exact expression
for $I(X,y)$ for uniform signal and noise and $\sigma\le 1$. In
addition,~\cite{Stocks.01a} finds a large $N$ approximation to the
mutual information. Using the same arguments as for the $\sigma=1$
case, this approximation can be improved to
\begin{equation}\label{5:InfouniformLargeN1}
    I(X,y)
    \simeq\frac{\sigma}{2}\log_2{\left(\frac{(N+2)e}{2\pi}\right)}+(1-\sigma)(1-\log_2{(1-\sigma)})-\sigma\log_2{(\sigma)}.
\end{equation}
The accuracy of Eq.~(\ref{5:InfouniformLargeN1}) can be verified by
numerical comparison with the exact formula in~\cite{Stocks.01a}, as
$N$ increases. If one replicates Fig.~3 of~\cite{Stocks.01a} in this
manner, it is clear that Eq.~(\ref{5:InfouniformLargeN1}) is the
more accurate approximation.

Differentiating Eq.~(\ref{5:InfouniformLargeN1}) with respect to
$\sigma$ and setting to zero obtains the optimal value of $\sigma$
as
\begin{equation}\label{5:maxsigUnif}
    \sigma_o =
    \frac{\sqrt{(N+2)}}{\sqrt{(N+2)}+\sqrt{\left(\frac{8\pi}{e}\right)}}.
\end{equation}
The channel capacity at $\sigma_o$ is
\begin{equation}\label{5:MaxInfoUnif}
    C(X,y) = 1-\log_2{(1-\sigma_o)} = \log_2{\left(2+\sqrt{\frac{(N+2)e}{2\pi}}\right)}.
\end{equation}
Clearly, $\lim_{N\rightarrow\infty}\sigma_o=1$, and the capacity
approaches $0.5\log_2{\left((N+2)e/(2\pi)\right)}$, which agrees
with Eq.~(\ref{5:InfoNew1}). Expressions for $\sigma_o$ and the
corresponding capacity for large $N$ are also given
in~\cite{Stocks.01a}. Again, these are slightly different to
Eqs.~(\ref{5:maxsigUnif}) and~(\ref{5:MaxInfoUnif}), due to the
slightly inaccurate terms in the large $N$ approximation to
$H(y|X)$. However the important qualitative result remains the same,
which is that the channel capacity scales with $0.5\log_2{(N)}$ and
the value of $\sigma$ which achieves this asymptotically approaches
unity.

\subsubsection{Gaussian Signal and Noise}

In~\cite{Hoch.03a}, an analytical approximation for $\sigma_o$ for
the specific case of Gaussian signal and noise is derived using a
Taylor expansion of the Fisher information inside the integral in
Eq.~(\ref{5:Info_largeN1}). We give a slightly different derivation
of this approach that uses the PDF $f_Q(\tau)$.

We begin with Problem~(\ref{5:ChannelCapacity2}). Solving this
problem requires differentiating $f(\sigma)$ with respect to
$\sigma$ and solving for zero. From Table~\ref{4:table3}, the
derivative of the relative entropy between $f_X$ and $f_\eta$ is
\begin{equation}\label{5:dRelEnt}
\frac{d}{d\sigma}D(f_X||f_\eta) =
\frac{1}{\ln{2}}\left(\sigma^{-1}-\sigma^{-3}\right).
\end{equation}
For the second term, $g(\sigma)$, we take the lead
from~\cite{Hoch.03a} and approximate $\ln{(P_{1|x})}$ by its second
order Taylor series expansion~\cite{Spiegel}. The result is that
\begin{equation}\label{5:Taylor2}
    g(\sigma) = -\int_{x=-\infty}^{x=\infty}f_X(x)\log_2{(P_{1|x})}dx \simeq
    1+\frac{1}{\pi\sigma^2\ln{2}}.
\end{equation}
Numerical testing finds that the approximation of
Eq.~(\ref{5:Taylor2}) appears to be quite accurate for all $\sigma$,
as the relative error is no more than about $10$ percent for $\sigma
> 0.2$. However, as we will see, this is inaccurate enough to cause
the end result for the approximate channel capacity to significantly
overstate the true channel capacity.

Taking the derivative of Eq.~(\ref{5:Taylor2}) with respect to
$\sigma$, subtracting it from Eq.~(\ref{5:dRelEnt}), setting the
result to zero and solving for  $\sigma$ gives the optimal value of
$\sigma$ found in~\cite{Hoch.03a}, $\sigma_o \simeq
\sqrt{1-\frac{2}{\pi}}\simeq 0.6028$.

An expression for the mutual information at $\sigma_o$ can be found
by back-substitution. Carrying this out gives the large $N$ channel
capacity for Gaussian signal and noise as
\begin{equation}\label{5:ChannelCapacity6}
    C(X,y) \simeq 0.5\log_2{\left(\frac{2N}{e(\pi-2)}\right)},
\end{equation}
which can be written as $C(X,y) \simeq 0.5\log_2{N} -0.3169$.

Although Eq.~(\ref{5:ChannelCapacity6}) is close to correct, recall
from Sec.~\ref{c:5capacity} that capacity must be less than
$0.5\log_2{N} -0.3956$ and hence Eq.~(\ref{5:ChannelCapacity6})
significantly overstates the true capacity.

\begin{acknowledgments}
This work was funded by the Australian Research Council, an
Australian Academy of Science Young Researcher's Award, as well as
EPSRC grant EP/C523334/1, and we gratefully acknowledge this
support.  The authors would also like to thank Priscilla Greenwood,
of Arizona State University, Pierre-Olivier Amblard of Laboratoire
des Images et des Signaux, France, Thinh Nguyen of Oregon State
University and Simon Durrant of the University of Plymouth, UK, for
valuable discussions and generous provision of preprints, and the
anonymous referees, whose comments have led to significant
improvements in this paper.
\end{acknowledgments}

\appendix

\section{Derivations}

\subsection{Mutual Information for Large $N$ and Arbitrary
$\sigma$}\label{c:5Fisher}

This appendix contains derivations of the large $N$ approximations
to the output entropy and mutual information discussed in
Sec.~\ref{c:5capacity}.

\subsubsection{Conditional Output Entropy}

An approximation to the conditional output entropy, $H(y|X)$, can be
derived by noting that for large $N$ the binomial distribution can
be approximated by a Gaussian distribution with the same mean and
variance---i.e. $NP_{1|x}$ and $NP_{1|x}(1-P_{1|x})$ respectively.
Provided $0{\ll}NP_{1|x}{\ll}N$ we have
\begin{equation}\label{5:pn_x_Gaussian}
    P_{y|X}(n|x) \simeq
    \frac{1}{\sqrt{2{\pi}NP_{1|x}(1-P_{1|x})}}\exp{\left(-\frac{(n-NP_{1|x})^2}{2NP_{1|x}(1-P_{1|x})}\right)}.
\end{equation}
The {\it average} conditional output entropy is $H(y|X) =
\int_xf_X(x)\hat{H}(y|x)dx$, where
\begin{equation}\label{5:hhaty_x}
    \hat{H}(y|x) = -\sum_{n=0}^NP_{y|X}(n|x)\log_2{(P_{y|X}(n|x))}.
\end{equation}
Using the well known result for the entropy of a Gaussian random
variable~\cite{Cover} we can write
\begin{align}\label{5:hhaty_x1}
    \hat{H}(y|x)&\simeq
    0.5\log_2{(2{\pi}eNP_{1|x}(1-P_{1|x}))}.
\end{align}
Multiplying both sides of Eq.~(\ref{5:hhaty_x1}) by $f_X(x)$ and
integrating over all $x$ gives
\begin{align}\label{5:Hyxnew}
    H(y|X)
    &\simeq0.5\log_2{(2{\pi}eN)}+0.5\int_{x=-\infty}^{\infty}f_X(x)\log_2{\left(P_{1|x}(1-P_{1|x})\right)}dx\notag\\
    &=0.5\log_2{(2{\pi}eN)}+0.5\int_{\tau=0}^{\tau=1}f_Q(\tau)\log_2{(\tau)}d\tau+\notag\\
    &0.5\int_{\tau=0}^{\tau=1}f_Q(\tau)\log_2{(1-\tau)}d\tau.
\end{align}
Eq.~(\ref{5:Hyxnew}) can be verified for the case where
$f_X(x)=f_\eta(\theta-x)$, since this means $f_Q(\tau)=1$ and
$\int_{\tau=0}^{\tau=1}f_Q(\tau)\log_2{(\tau)}d\tau=-\log_2{(e)}$.
Consequently Eq.~(\ref{5:Hyxnew}) reduces to $H(y|X) \simeq
0.5\log_2{\left(\frac{2{\pi}N}{e}\right)}$ which agrees precisely
with Eq.~(\ref{5:Hy_xNew1}). This approximation breaks down when
$P_{1|x}$ is close to zero or unity. Furthermore,
Eq.~(\ref{5:Hyxnew}) holds exactly only for values of $x$ for which
$P_{y|X}(n|x)$ is exactly Gaussian. Otherwise, $H(y|X)$ is strictly
less than the approximation given.

\subsubsection{Output Distribution and entropy}\label{Laplace}

For large $N$, since $P_{y|X}(n|x)$ is Gaussian, $y/N$ approaches a
delta function located at $P_{1|x}=n/N$. From Eqs.~(\ref{4:Py})
and~(\ref{4:pstar_n}), this means that $P_y(n)$ can be written in
terms of the PDF of the average transfer function, $f_Q(\cdot)$, as
\begin{equation}\label{5:Pyn_Q1}
    P_y(n) \simeq \frac{f_Q\left(\frac{n}{N}\right)}{N}.
\end{equation}
This result can be derived more rigorously using saddlepoint
methods~\cite{Brunel.98}.

\begin{figure}[hbtp]
\begin{center}
{\subfigure[~$\sigma=0.4$, Gaussian signal and
noise]{\includegraphics[scale=0.4]{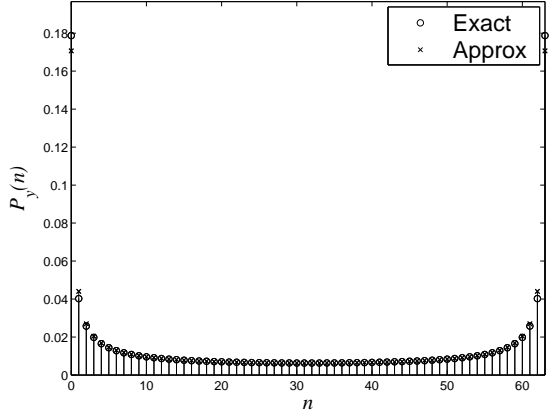}\label{f:5:PyLargeN1G}}
~ \subfigure[~$\sigma=1.6$, Gaussian signal and
noise]{\includegraphics[scale=0.4]{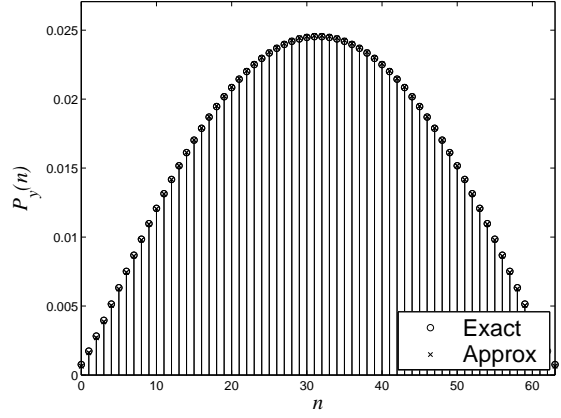}\label{f:5:PyLargeN3G}}
} \caption[Large $N$ approximation to $P_y(n)$]{Approximation to the
output PMF, $P_y(n)$, given by Eq.~(\ref{5:Py4}), for $N=63$.
Circles indicate the exact $P_y(n)$ obtained by numerical
integration and the crosses show
approximations.}\label{f:5:PyLargeNG}
\end{center}
\end{figure}

Consider the case where the signal and noise both have the same
distribution but different variances. When the noise intensity,
$\sigma
> 1$, then $f_Q(0)=f_Q(1)=0$, whereas for $\sigma <1$, we have $f_Q(0)=f_Q(1)=\infty$. From
Eq.~(\ref{5:Pyn_Q1}), this means $P_y(0)$ and $P_y(N)$ are either
zero or infinite. However, for finite $N$, there is some finite
nonzero probability that all output states are on or off. Indeed, at
$\sigma=1$, we know that $P_y(n)=\frac{1}{N+1}~\forall~n$, and at
$\sigma=0$, $P_y(0)=P_y(N) = 0.5$. Furthermore, for finite $N$,
Eq.~(\ref{5:Pyn_Q1}) does not guarantee that $\sum_{n=0}^NP_y(n)=1$.
To increase the accuracy of our approximation by ensuring $P_y(0)$
and $P_y(N)$ are always finite, and that $P_y(n)$ forms a valid PMF,
we define a new approximation as
\begin{equation}\label{5:Py4}
    P_y'(n) = \left\{ \begin{array}{ll}
   \frac{f_Q\left(\frac{n}{N}\right)}{N} & \quad\mbox{for}\quad n=1,..,N-1\\
    0.5\left(1-\sum_{m=1}^{N-1}\frac{f_Q\left(\frac{m}{N}\right)}{N}\right) & \quad\mbox{for}\quad
    n=0,n=N.
    \end{array}\right.
\end{equation}
Fig.~\ref{f:5:PyLargeNG} shows that the approximation given by
$P_y'(n)$ is highly accurate for $N$ as small as $63$, for $\sigma$
both smaller and larger than unity.

Consider the entropy of the discrete random variable $y$. Making use
of Eq.~(\ref{5:Pyn_Q1}), we have
\begin{align}\label{5:Hy_largeN1}
    H(y) &=-\sum_{n=0}^{N}P_y(n)\log_2{(P_y(n))}\notag\\
    &=-\frac{1}{N}\sum_{n=0}^{N}f_Q\left(\frac{n}{N}\right)\log_2{\left(f_Q\left(\frac{n}{N}\right)\right)}+\notag\\
    &\frac{\log_2{(N)}}{N}\sum_{n=0}^Nf_Q\left(\frac{n}{N}\right).
\end{align}

Suppose that the summations above can be approximated by integrals,
without any remainder terms. Carrying this out and then making the
change of variable $\tau = n/N$ gives
\begin{align}\label{5:Hy_largeN2}
    H(y) &\simeq\log_2{N}
    -\int_{\tau=0}^{\tau=1}f_Q(\tau)\log_2{\left(f_Q(\tau)\right)}d\tau\notag\\
    &=\log_2{N} + H(Q),
\end{align}
where $H(Q)$ is the differential entropy of the random variable $Q$.
Performing a change of variable in Eq.~(\ref{5:Hy_largeN2}) of
$\tau=1-F_\eta(\theta-x)$ gives
\begin{equation}\label{5:Hy_largeN3}
    H(y) \simeq
    \log_2{(N)}-D(f_X(x)||f_\eta(\theta-x)).
\end{equation}
This result shows that $H(y)$ for large $N$ is approximately the sum
of the number of output bits and the negative of the relative
entropy between $f_X$ and $f_\eta$. Therefore, since relative
entropy is always non-negative, the approximation to $H(y)$ given by
Eq.~(\ref{5:Hy_largeN3}) is always less than or equal to
$\log_2{(N)}$. This agrees with the known expression for $H(y)$ in
the specific case of $\sigma=1$ of $\log_2{(N+1})$, which holds for
any $N$.

\subsubsection{Mutual Information}\label{c:5Info}

Subtracting Eq.~(\ref{5:Hyxnew}) from Eq.~(\ref{5:Hy_largeN2}) gives
a large $N$ approximation to the mutual information as
\begin{align}%\label{5:Info_largeN}
    I(X,y) &\simeq 0.5\log_2{\left(\frac{N}{2{\pi}e}\right)}+H(Q)\notag\\
    &-0.5\int_{\tau=0}^{\tau=1}f_Q(\tau)\log_2{\left({\tau(1-\tau)}\right)}d\tau.\label{5:Info_largeN_3}
\end{align}
As discussed in the main text, the mutual information scales with
$0.5\log_2{(N)}$. The importance of the $N$-independent terms in
Eq.~(\ref{5:Info_largeN_3}) is that they determine how the mutual
information varies from $0.5\log_2{\left(\frac{N}{2{\pi}e}\right)}$
for different PDFs, $f_Q(\tau)$.

\begin{figure}[hbtp]
\begin{center}
{\subfigure[~Gaussian]{\includegraphics[scale=0.4]{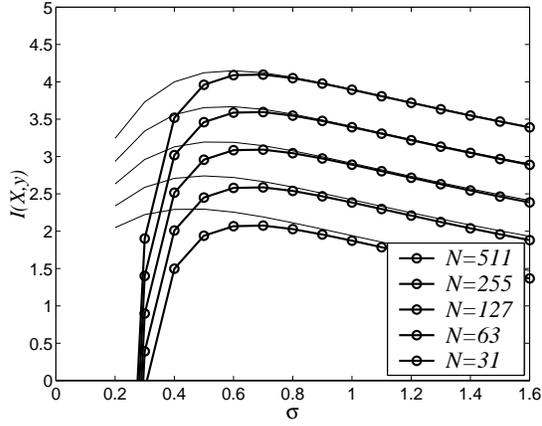}\label{f:5:InfoLargeNGaussian_sub}}
~
\subfigure[~Laplacian]{\includegraphics[scale=0.4]{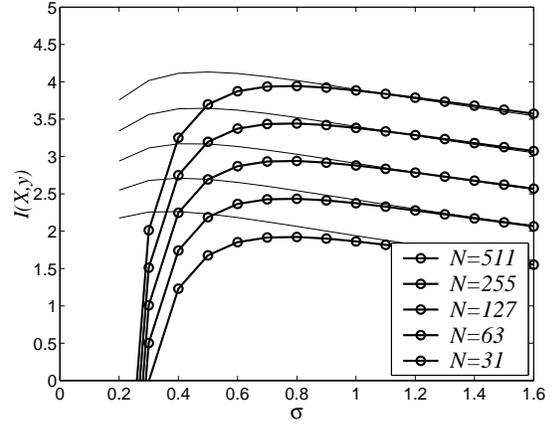}\label{f:5:InfoLargeNLaplacian_sub}}}
\caption{Large $N$ approximation to mutual information given by
Eq.~(\ref{5:Info_largeN_3}) and exact mutual information calculated
numerically. The exact expression is shown by thin solid lines, and
the approximation by circles, with a thicker solid line
interpolating between values of $\sigma$ as an aid to the eye. The
approximation can be seen to always be a lower bound on the exact
mutual information.}\label{f:5:InfoLargeN}
\end{center}
\end{figure}

Fig.~\ref{f:5:InfoLargeN} shows, as examples, the approximation of
Eq.~(\ref{5:Info_largeN_3}), as well as the exact mutual
information---calculated by numerical integration---for the Gaussian
and Laplacian cases, for a range of $\sigma$ and increasing $N$. As
with the output entropy, the mutual information approximation is
quite good for $\sigma
> 0.7$, but worsens for smaller $\sigma$. However, as $N$
increases the approximation improves.

Eq.~(\ref{5:Info_largeN_3}) can be rewritten via the change of
variable, $x=\theta-F_\eta^{-1}(1-\tau)$, as
\begin{align}\label{5:Info_largeNa}
    I(X,y)
    &=0.5\log_2{\left(\frac{N}{2{\pi}e}\right)}-\notag\\
    &\int_{x=-\infty}^{x=\infty}f_X(x)\log_2{(P_{1|x}(1-P_{1|x}))}dx-D(f_X(x)||f_\eta(\theta-x)).
\end{align}
Rearranging Eq.~(\ref{5:Info_largeNa}) gives
Eq.~(\ref{5:Info_largeN1})---with the Fisher information, $J(x)$,
given by Eq.~(\ref{5:Fisher_I})---which is precisely the same as
that derived in~\cite{Hoch.03a} as an asymptotic large $N$
expression for the mutual information. Our analysis
extends~\cite{Hoch.03a} by finding large $N$ approximations to both
$H(y)$ and $H(y|X)$, as well as the output distribution, $P_y(n)$.
We have also illustrated the role of the PDF, $f_Q(\tau)$, in these
approximations, and justified the use of Eq.~(\ref{5:Info_largeN1})
for the SSR model.

\subsection{Proof that $f_S(x)$ is a PDF}\label{c:A5Jeffrey}

As shown in~\cite{Hoch.03a}, the Fisher information for the SSR
model is given by Eq.~(\ref{5:Fisher_I}). Consider $f_S(x)$ as in
Eq.~(\ref{f_S}). Since $f_\eta(x)$ is a PDF and $F_\eta(x)$ is the
CDF of $\eta$ evaluated at $x$, we have $f_S(x) \ge 0~\forall~x$.
Letting $h(x)=F_\eta(\theta-x)$, Eq.~(\ref{f_S}) can be written as
\begin{equation}\label{A5:sqrtJ3}
    f_S(x) = \frac{-h'(x)}{\pi\sqrt{h(x)-h(x)^2}}.
\end{equation}
Suppose $f_\eta(x)$ has support $x\in[-a,a]$. Integrating $f_S(x)$
over all $x$ gives
\begin{align}\label{A5:sx}
    \int_{x=-a}^{x=a}f_S(x)dx &=
    \int_{x=-a}^{x=a}\frac{-h'(x)}{\pi\sqrt{h(x)-h(x)^2}}dx\notag\\
    &= -\frac{1}{\pi}\left(2\arcsin{\left(\sqrt{h(x)}\right)}|_{x=-a}^{x=a}\right)\notag\\
    &= -\frac{2}{\pi}\left(\arcsin(0)-\arcsin(1)\right)=1,
\end{align}
which means $f_S(x)$ is a PDF.

\subsection{$H(y|X)$ for large $N$ and
$\sigma=1$}\label{A:Hy_x}

Here we derive a large $N$ approximation to $H(y|X)$  used in
Sec.~\ref{c:5sig1}. For $\sigma=1$ the output PMF is $P_y(n) =
\frac{1}{N+1}$ $\forall$ $n$~\cite{Stocks.01b}. Using this, it can
be shown that
\begin{equation}\label{5:sumlog}
    -\sum_{n=0}^NP_y(n)\log_2{{N\choose n}} = \log_2{(N!)}-\frac{2}{N+1}\sum_{n=1}^Nn\log_2{n}.
\end{equation}
We will now see that both terms of Eq.~(\ref{5:sumlog}) can be
simplified by approximations that hold for large $N$. Firstly, for
the $\log_2{(N!)}$ term, we can make use of Stirling's
formula~\cite{Spiegel}, which is valid for large $N$,
\begin{equation}\label{5:Stirling}
    N! \sim \sqrt{(2{\pi}N)}N^N\exp{(-N)}.
\end{equation}
This approximation is particularly accurate if the log is taken of
both sides, which we require. Secondly, the sum in the second term
of Eq.~(\ref{5:sumlog}) can be approximated by an integral and
simplified by way of the Euler-Maclaurin summation
formula~\cite{Spiegel}. The result is
\begin{equation}\label{5:Emac5}
    \frac{2}{N+1}\sum_{n=1}^Nn\log_2{n} \simeq
    N\log_2{(N+1)}-\frac{N(N+2)}{2\ln{2}(N+1)}+O\left(\frac{\log{N}}{N}\right).
\end{equation}
Subtracting Eq.~(\ref{5:Emac5}) from the log of
Eq.~(\ref{5:Stirling}) gives
\begin{align}\label{5:Emac6}
-\sum_{n=0}^NP_y(n)\log_2{{N\choose n}} \simeq&~
 0.5\log_2{\left(\frac{N}{e^2}\right)}-\frac{N}{2\ln{2}}\left(2-\frac{N+2}{N+1}\right)\notag\\
&+0.5\log_2{(2\pi)}-O\left(\frac{\log{N}}{N}\right),
\end{align}
where we have used
$N\log_2{(1+\frac{1}{N})}=\frac{1}{\ln{2}}+O\left(\frac{1}{N}\right)$.
When Eq.~(\ref{5:Emac6}) is substituted into an exact expression for
$H(y|X)$ given in~\cite{Stocks.01b}, we get
\begin{align}\label{5A:Hy_xNew}
    H(y|X) =&~ \frac{N}{2\ln{2}}-\sum_{n=0}^NP_y(n)\log_2{{N\choose
    n}}\notag\\
     \simeq&~0.5\log_2{N}+0.5\left(\frac{N}{N+1}-2\right)\log_2{(e)}+\notag\\
     &0.5\log_2{(2\pi)}-O\left(\frac{\log{N}}{N}\right).
\end{align}
The final result is that for large $N$,
\begin{equation}\label{5:Hy_xNew1}
    H(y|X)
    \simeq 0.5\log_2{\left(\frac{2{\pi}N}{e}\right)}.
\end{equation}

\clearpage

\clearpage

\clearpage

\clearpage

\clearpage

\clearpage

\begin{table}[t]
\centering \caption[]{The auxiliary PDF, $f_Q(\tau)$, for five
different `matched' signal and noise distributions (i.e. same
distribution but with different variances), as well as $H(Q)$, the
entropy of $f_Q(\tau)$. The threshold value, $\theta$, and the
signal and noise means are assumed to be zero, so that these results
are independent of $\theta$. The noise intensity,
$\sigma=\sigma_\eta/\sigma_x$, is the ratio of the noise standard
deviation to the signal standard deviation. For the Cauchy case,
$\sigma_\lambda$ is the ratio of the full-width-at-half-maximum
parameters. The label `NAS' indicates that there is no analytical
solution for the entropy.}\label{4:table3}
\begin{tabular}{|c|c|c|}
  % after \\: \hline or \cline{col1-col2} \cline{col3-col4} ...
\hline
  Distribution& $f_Q(\tau)$ &$H(Q)$\\
  \hline
  Gaussian & $\sigma\exp{\left((1-\sigma^2)\left(\mbox{erf}^{-1}(2\tau-1)\right)^2\right)}$& $-\log_2{(\sigma)}-\frac{1}{2\ln{2}}\left(\frac{1}{\sigma^2}-1\right) $\\
  Uniform, $\sigma \ge 1$  & $\left\{\begin{array}{rll}
    &\sigma, & -\frac{1}{2\sigma}+0.5\le \tau \le \frac{1}{2\sigma}+0.5,\\
    &0, & \mbox{ otherwise.}\end{array}\right.$ & $\log_2{\sigma}$\\
  Laplacian& $\left\{ \begin{array}{rll}
        &\sigma(2\tau)^{(\sigma-1)} &\mbox{ for } 0\le \tau\le 0.5,\\
        & \sigma(2(1-\tau))^{(\sigma-1)}&\mbox{ for } 0.5 \le \tau \le 1.
    \end{array}\right.$ &$-\log_2{(\sigma)}-\frac{1}{2\ln{2}}\left(\frac{1}{\sigma}-1\right)$ \\
  Logistic & $\sigma\frac{(\tau(1-\tau))^{(\sigma-1)}}{\left(\tau^\sigma+(1-\tau)^\sigma\right)^2}$ &\mbox{NAS}\\
  Cauchy & $\sigma_\lambda\frac{1+\tan^2{(\pi(\tau-0.5))}}{(1+\sigma_\lambda^2\tan^2{(\pi(\tau-0.5))})}$ &\mbox{NAS}\\
  \hline
 \end{tabular}
\end{table}

\clearpage

\begin{table}[hbtp]
\centering \caption[]{Large $N$ channel capacity and optimal
$\sigma$ for `matched' signal and
noise}\label{5:table1}\vspace{0.2cm}
\begin{tabular}{|c|c|c|c|}
  % after \\: \hline or \cline{col1-col2} \cline{col3-col4} ...
\hline
  Distribution& $C(X,y)-0.5\log_2{(N)}$ &$\sigma_o$& $C(X,y)-I(X,y)|_{\sigma=1}$\\
  \hline
  Gaussian & $-0.3964$& $0.6563$ &$0.208$\\
  Logistic& $-0.3996$ &$0.5943$ & $0.205$\\
  Laplacian & $-0.3990$ &$0.5384$ & $0.205$\\
  \hline
 \end{tabular}
\end{table}

\end{document}